\documentclass{article}
\usepackage{spconf,amsmath,graphicx}
\usepackage[hidelinks]{hyperref}
\usepackage[sort]{cite}
\usepackage{tikz}
\usetikzlibrary{arrows.meta,calc}
\usepackage{booktabs} 
\usepackage{adjustbox} 
\usepackage{amssymb}  
\usepackage{makecell}
\usepackage{diagbox}
\usepackage{siunitx}
\usepackage{multirow} 
\usepackage{enumitem}
\usepackage{bm}  
\usepackage{pbalance}

\usepackage{acronym}
\acrodef{DNN}{deep neural network}
\acrodef{GAN}{generative adversarial network}
\acrodef{DoA}{direction of arrival}
\acrodef{PDF}{probability density function}
\acrodef{E2E}{end-to-end}
\acrodef{STFT}{short-time Fourier transform }
\acrodef{TF}{time-frequency}
\acrodef{RIR}{room impulse response}
\acrodef{FiLM}{feature-wise linear modulation }
\acrodef{SNR}{signal-to-noise ratio}
\acrodef{MOS}{mean opinion score}
\acrodef{DNN}{deep neural network}
\acrodef{PDF}{probability density function}
\acrodef{FiLM}{feature-wise linear modulation }
\acrodef{TSE}{Target speaker extraction }
\acrodef{SE}{speech enhancement}
\acrodef{SOTA}{state-of-the-art}
\acrodef{SDE}{stochastic differential equation}
\acrodef{ODE}{ordinary differential equation}
\acrodef{WER}{word error rate}
\acrodef{CER}{character error rate}
\acrodef{CFM}{conditional flow matching}

\newcommand{\xv}{\mathbf{x}}
\newcommand{\yv}{\mathbf{y}}

\title{A Comparison of Generative and Discriminative Methods for Speech Enhancement: Robustness, Complexity, and Hallucination}
\name{Shrishti Saha Shetu$^{1}$, Emanu\"{e}l A. P. Habets$^{1}$,  Andreas Brendel $^{2}$}

\address{
$^{1}$International Audio Laboratories Erlangen\footnotemark[1], Am Wolfsmantel 33, 91058 Erlangen, Germany\\
$^{2}$Fraunhofer IIS, Am Wolfsmantel 33, 91058 Erlangen, Germany\\
{\small \{shrishti.saha.shetu, emanuel.habets, andreas.brendel\}@iis.fraunhofer.de}
}

\begin{document}
\ninept
\maketitle

\footnotetext[1]{A joint institution of Fraunhofer IIS and Friedrich-Alexander-Universit{\"a}t Erlangen-N{\"u}rnberg (FAU), Germany.}

\begin{abstract}
In this study, we conduct a comprehensive comparative analysis of generative and discriminative deep learning-based speech enhancement methods, specifically in noise reduction tasks. Our investigation focuses on evaluating their effectiveness under high and low signal-to-noise ratio conditions, considering both matched and mismatched training scenarios. We further investigate the impact of training data volume, model convergence speed, and interpret the performance differences in terms of objective results for the considered training paradigms. Additionally, we compare the complexity-performance trade-off and the practical viability of these approaches. To further strengthen the evaluation, we study the hallucination characteristics of generative approaches in terms of word error rate and phoneme similarity. The insights derived from this study provide empirical evidence to assist researchers and practitioners in understanding  whether the perceptual gains of different approaches justify their computational cost in practical applications.

\end{abstract}
\section{Introduction}
\label{sec:intro}
Research in the \ac{SE} domain has experienced considerable progress in recent years, largely attributed to \ac{DNN}-based approaches that promise significant improvements in speech quality and intelligibility even under challenging acoustic conditions. Although discriminative \ac{DNN}-based methods remain widely used and have already been deployed in numerous practical applications on consumer devices, researchers in the \ac{SE} domain are increasingly investigating generative methods for this task \cite{hu2020dccrn, choi2021real, shetu2023ultra, tan2019learning, fu2021metricgan+, ramonaite2026generative, richter2023speech, zhao2025conditional}. This shift is largely motivated by the need to mitigate speech distortions introduced in many discriminative methods and to achieve superior performance in very low-SNR scenarios \cite{wang2019bridging, shetu2025leveraging}. 

Generative methods, such as \ac{GAN}- and diffusion-type approaches have been shown to achieve \ac{SOTA} performance in various \ac{SE} tasks \cite{scheibler24_interspeech, gonzalez2024investigating, shetu2025gan}. Although showing promising performance, the majority of generative methods in the literature rely on very large \ac{DNN} models, which impose significant computational constraints for practical deployment. Moreover, most \ac{CFM}- and diffusion-based approaches rely on an iterative generation process, further increasing the overall computational complexity. As a result, it becomes challenging to systematically evaluate the perceptual performance gains in relation to the associated computational cost when comparing discriminative and generative methods. Such a comparison requires training and evaluating a large set of models while carefully evaluating the trade-offs between performance and computational complexity. As another critical aspect, many existing approaches are trained and evaluated on small and homogeneous datasets (e.g., VCTK dataset \cite{valentini2016investigating}, EARS dataset \cite{richter2024ears} with WHAM noises \cite{Wichern2019WHAM}) that lack diversity in noise types and \ac{SNR} conditions; factors that are common in real-world applications \cite{richter2023speech, cao2022cmgan, fu2019metricgan}. We argue that properly assessing the true benefits of these different approaches requires a comprehensive comparative study that considers multiple evaluation aspects.

In \cite{shetu2024comparative}, the performance of different \ac{SOTA} discriminative methods has been evaluated in low \ac{SNR} scenarios. Several studies also investigate the impact of different training objectives for \ac{SE} \cite{10095258, 10887784}. However, to the best of our knowledge, no study in the literature comprehensively considers a wide range of discriminative and generative methods, diverse model architectures, and varied training and evaluation datasets. Furthermore, existing studies do not jointly evaluate performance using both classical and \ac{DNN}-based metrics alongside the complexity–performance trade-off. In this work, we present a comprehensive evaluation of these methods for noise reduction tasks. Our specific contributions are as follows:

\begin{itemize}[leftmargin=0pt, labelindent=0pt]
    \item We trained a large set of discriminative and generative models on diverse datasets to evaluate their performance under various noise conditions, including low-\ac{SNR} and mismatched training scenarios.
    \item We analyze the convergence characteristics of different methods and investigate the impact of training data volume on performance, as well as the hallucination tendencies of these methods.
    \item We examine the complexity–performance trade-off between different approaches and provide insights into when each method is most suitable for practical use.
\end{itemize}

\section{Methods}
\label{sec:PM}

Let \( p(\xv_0 \mid \yv) \) denote the conditional distribution of clean speech \( \xv_0 \in \mathbb{R}^L\) given a noisy signal \( \yv \in \mathbb{R}^L\), where $L$ is the length in samples in the time domain. In conditional generative \ac{SE}, the objective is to learn a transformation of samples from a source distribution such that the transformed samples follow the target clean speech distribution using information from \( \yv \). For brevity, we assume that all transformations preserve the signal dimensions and, hence, do not mention their dimensions in the remainder of this paper.

\subsection{Diffusion-Type Models}

\textbf{Score Matching:} Diffusion models~\cite{sohl2015deep, ho2020denoising} typically define a stochastic transform from clean data into Gaussian noise via a forward \ac{SDE}.  In the \ac{SE} context (e.g., SGMSE+~\cite{richter2023speech}), the diffusion kernel $q(\xv_t \mid \xv_0, \yv)$ is used to perturb the clean speech signal toward the noisy signal $\yv$ for different time steps $t \in [0, T]$. Here, $\xv_t$ denotes the noisy signal at timestep $t$ according to an additive Gaussian noise schedule along the path to $\yv$ defined by the forward \ac{SDE}~\cite{richter2023speech}. A score network $\mathbf{S}_\theta(\xv_t, \yv, t)$, with parameters $\theta$, is trained to approximate the conditional score function $\nabla_{\xv_t} \log p_t(\xv_t \mid \yv)$ by minimizing the denoising score-matching objective
\begin{equation}
\mathcal{L}_{\mathrm{diff}} =
\mathbb{E}_{t, \mathbf{x}_0, \mathbf{y}, \boldsymbol{\epsilon}}
\left[
\left\| \boldsymbol{\epsilon} - \mathbf{S}_\theta(\mathbf{x}_t, \mathbf{y}, t) \right\|^2_2
\right],
\quad \boldsymbol{\epsilon} \sim \mathcal{N}(\mathbf{0}, \mathbf{I}).
\label{eq:loss_diffusion}
\end{equation}
Here, $\boldsymbol{\epsilon}$ denotes the diffusion noise at diffusion time step $t$, where $t$ is sampled from a uniform distribution $t \sim \mathcal{U}(0, T)$. While the network $\mathbf{S}_\theta$ estimates diffusion noise $\boldsymbol{\epsilon}$, the loss \eqref{eq:loss_diffusion} is mathematically equivalent to score function estimation up to a time-dependent scaling factor.   By solving the reverse-time \ac{SDE}, the model iteratively refines the noisy observation into a clean speech estimate. In this work, we consider SGMSE+~\cite{richter2023speech} and BBED~\cite{lay2023reducing} as diffusion models. 
In addition, the score-matching objective can be reformulated by directly predicting the initial clean speech signal $\mathbf{x}_0$ at each diffusion step $t$~\cite{richter2023speech}. For this formulation, we consider GALDSE~\cite{wang2024gald}.

\noindent\textbf{Conditional Flow Matching (CFM):} \ac{CFM} frameworks, such as FlowSE~\cite{lee2025flowse}, model \ac{SE} via sample transport by a (deterministic) velocity field $\mathbf{v}_\theta(\xv_t, \yv, t)$. These approaches define a probability path $\{p_t(\xv)\}_{t \in [0,1]}$ connecting the noisy signal $\yv$ (at $t=1$) to the clean target $\xv_0$ (at $t=0$). The model of the velocity field is trained to match a conditional target vector field $\mathbf{u}_t(\xv_t \mid \xv_0, \yv)$
\begin{equation}
\mathcal{L}_{\mathrm{CFM}} =
\mathbb{E}_{t, \xv_0, \yv}\left[\left\|\mathbf{v}_\theta(\xv_t, \yv, t)
-\mathbf{u}_t(\xv_t \mid \xv_0, \yv)\right\|^2_2\right].
\end{equation}
For optimal transport (OT) paths, $\mathbf{u}_t$ reduces to the constant velocity $\xv_0 - \yv$, enabling efficient deterministic transport along a straight line via an \ac{ODE}. In this work, we consider FlowSE~\cite{lee2025flowse} for this formulation.

\noindent\textbf{Consistency Models:} Consistency models~\cite{song2023consistency} enforce a trajectory-invariance property to eliminate the need for iterative refinement in \ac{SDE} or \ac{ODE} solvers. A consistency function $f_\theta(\xv_t, \yv, t)$ is trained to map any point $\xv_t$ on a trajectory directly to its origin $\xv_0$. The objective enforces self-consistency across all time steps $t \in [0, T]$
\begin{equation}
\mathcal{L}_{\mathrm{cons}} =
\mathbb{E}_{n, \xv_0, \yv}
\left[\left\|f_\theta(\xv_{t_n}, \yv, t_{n})-f_{\theta^-}(\xv_{t_{n-1}}, \yv, t_{n-1})
\right\|^2_2\right].
\end{equation}
where $n$ is sampled uniformly from $\{1, \dots, N\}$, with $N$ denoting the total number of discrete time steps, $\xv_{t_{n}}$ and $\xv_{t_{n-1}}$ denote adjacent points on the trajectory and $\theta^-$ is obtained by an exponential moving average (EMA) of the parameters $\theta$. This approach enables single-step \ac{SE}. As a representative method, we consider SEBridge~\cite{qiu2023se}.

\subsection{Generative Adversarial Networks (GANs)}
Conditional \ac{GAN}s, such as DisCoGAN and NoCoGAN (where only the noisy signal $\yv$ is used for conditioning)~\cite{shetu2025leveraging}, learn a mapping $G_\theta(\yv)$ that transforms a noisy signal $\yv$ directly into an estimate of its clean version. In this framework, the generator $G_\theta$ is trained through a min-max optimization with a discriminator $D_\phi$. The training objective is defined as
\begin{equation}
\min_{\theta} \max_{\phi} \; \mathbb{E}_{\xv_0, \yv }
\big[ \mathcal{L}_D(\phi, \xv_0, \yv) \big]+\mathbb{E}_{\yv}\big[ \mathcal{L}_G(\phi, \theta, \yv) \big].
\end{equation}
Here, $\mathcal{L}_G$ and $\mathcal{L}_D$ denote the generator and discriminator loss functions, respectively.  This approach enables direct single-step \ac{SE} without relying on any iterative generation process. In this work, we trained NoCoGAN and DisCoGAN~\cite{shetu2025leveraging}, as well as CMGAN~\cite{cao2022cmgan}. We also train the backbone network NCSN++ \cite{song2020score}, which is used in previously discussed diffusion-type methods, with a \ac{GAN} objective as described in \cite{shetu2025gan}; we refer to this variant as NCSN++ (GAN).

\subsection{Discriminative Models}
In contrast to generative paradigms, discriminative models approach \ac{SE} as a regression problem. These models focus on learning a deterministic mapping $\xv_0 \approx \mathcal{F}_\theta(\yv)$ by minimizing a signal- or mask-level loss function $\ell$ (e.g., SI-SNR or mean-squared error)
\begin{equation}
\mathcal{L}_{\mathrm{disc}} =
\mathbb{E}_{\xv_0, \yv}
\left[
\ell\!\left(\xv_0, \mathcal{F}_\theta(\yv)\right)
\right].
\end{equation}
In this work, we consider DCCRN~\cite{hu2020dccrn} and GCRN~\cite{tan2019learning}. We also train NoCoGAN and NCSN++ discriminatively using the reconstruction loss described in \cite{du2023funcodecfundamentalreproducibleintegrable}; we refer to these variants as NoCoGAN (D) and NCSN++ (D).

\section{Experiments}
\label{ED}

\subsection{Implementation details}
\noindent \textbf{Training and Evaluation Datasets:} We created training data using the Interspeech 2020 DNS Challenge dataset~\cite{reddy2020interspeech} by mixing clean speech with noise at random \ac{SNR}s within $[-25, 0]\, \mathrm{dB}$ for a low-\ac{SNR} dataset and within $[-5, 30]\, \mathrm{dB}$ for a high-\ac{SNR} dataset, each comprising approximately $1000$ hours of data, following \cite{shetu2024comparative,shetu2025leveraging}. 


For evaluation, we use the DNS Challenge non-reverberant test set~\cite{reddy2020interspeech} for high-\ac{SNR} scenarios, which contains noises from $12$ VoIP-relevant categories at \ac{SNR}s in $[0, 25]\, \mathrm{dB}$. For low-\ac{SNR} scenarios, we use the dataset from~\cite{shetu2025leveraging}, which comprises $1200$ samples, each lasting $10$~s. The dataset is divided into four \ac{SNR} groups, $[-15, -12]$, $[-11, -8]$, $[-7, -4]$, and $[-3, 0]\, \mathrm{dB}$, and contains $20$ different stationary and non-stationary noise types.

\begin{table}[t]
\caption{Results on the DNS Challenge non-reverberant test set. All reported models are trained on the high-\ac{SNR} dataset, representing matched conditions. The best results in each category are bold and overall best results are further underlined.}
\footnotesize
\centering
\setlength{\tabcolsep}{2.5pt}
\begin{tabular}{
    l
    l
    S[table-format=2.2]
    S[table-format=1.2]
    S[table-format=1.2]
    S[table-format=1.2]
}
\toprule
\cmidrule(lr){3-4} \cmidrule(lr){5-6}
\textbf{Method} & \textbf{Model}  
& {\makecell{\textbf{SI-SDR} \\ \textbf{(↑)}}} & {\makecell{\textbf{PESQ} \\ \textbf{(↑)}}}
& {\makecell{\textbf{SCOREQ} \\ \textbf{(↓)}}}& {\makecell{\textbf{DNSMOS} \\ \textbf{(↑)}}} \\
\midrule
\multirow{2}{*}{Unproc.}
& Noisy     & 9.06 & 1.58 & {0.93} & {3.15}  \\
& Clean    & {-} & {-} & {0} & {4.01}  \\
\midrule
\multirow{4}{*}{Disc.}
& DCCRN      & 17.36 & 2.91&  0.31 & 4.00  \\
& GCRN       & 16.71 & 2.63 & 0.42 & 3.91 \\
& NoCoGAN (D)  & 17.72 & \textbf{3.15} & 0.29 & 3.98 \\
& NCSN++ (D)    & \textbf{17.99} & 3.04 &  \textbf{ 0.28} & \textbf{4.02} \\
\midrule
\multirow{6}{*}{Diff.}
& SGMSE+    & 16.86 & 2.81 & 0.29 & 4.01 \\
& BBED     & \textbf{19.10} & \textbf{2.81} & \textbf{0.26} & 4.11 \\
& GALDSE   & 18.04 & 2.77 & 0.30 & \underline{\textbf{4.15}} \\
& SEBridge & 17.21 & 2.45 & 0.38& 4.00 \\
& SToRM    & 17.56 & 2.80 & 0.27 & 4.11 \\
& FlowSE & 17.90 & 2.73 & 0.28& 4.11 \\
\midrule
\multirow{4}{*}{\textbf{GAN}}
& NoCoGAN   & 17.82 & 3.22 & 0.29& 4.04 \\
& DisCoGAN  & 18.74 & \underline{\textbf{3.30}} & 0.25& 4.08 \\
& CMGAN     & 17.68& 2.93 & 0.36 & 4.02 \\
& NCSN++ (GAN)& \underline{\textbf{19.13}}& 3.19 & \underline{\textbf{0.24}} & \textbf{4.11} \\
\bottomrule
\end{tabular}
\label{tab:DNSnoreverb}
\vspace{-1.5em}
\end{table}

\begin{table*}[!t]
\caption{Evaluation of different discriminative and generative models trained on the low~\ac{SNR} dataset and evaluated in low-\ac{SNR} scenarios (matched condition). The best results in each category and for each \ac{SNR} group are bold and overall best results are further underlined.}
\small
\centering
\resizebox{\textwidth}{!}{%
\begin{tabular}{l l c c c c c c c c c c c c}
\toprule
\multicolumn{2}{l}{} 
& \multicolumn{4}{c}{\textbf{$\Delta$PESQ}(↑)} 
& \multicolumn{4}{c}{\makecell{\textbf{$\Delta$SI-SDR (dB)}(↑)}} 
& \multicolumn{4}{c}{\textbf{$\Delta$FwSegSNR}(↑)} \\
\cmidrule(lr){3-6} 
\cmidrule(lr){7-10} 
\cmidrule(lr){11-14}

\textbf{Method} & \textbf{Model} 
& \makecell{[-15,-12]} & \makecell{[-11,-8]} & \makecell{[-7,-4]} & \makecell{[-3,0]} 
& \makecell{[-15,-12]} & \makecell{[-11,-8]} & \makecell{[-7,-4]} & \makecell{[-3,0]} 
& \makecell{[-15,-12]} & \makecell{[-11,-8]} & \makecell{[-7,-4]} & \makecell{[-3,0]} \\
\midrule

\multirow{4}{*}{Disc.}
& GCRN & 0.33 & 0.44 & 0.58 & 0.71 & 15.34 & 14.39 & 13.04 & 11.31 & 1.84 & 2.53 & 3.31 & 3.65 \\
& DCCRN & 0.40 & 0.55 & 0.71 & 0.88 & 16.19 & 14.94 & 13.57 & 11.73 & 2.59 & 3.52 & 4.17 & 4.64 \\
& NoCoGAN (D) & 0.54 & \textbf{0.73} & \textbf{0.93} & \textbf{1.10} & 16.70 & 15.20 & 13.75 & 11.84 & 5.85 & 6.91 & 7.39 & 7.86 \\
& NCSN++ (D) & \textbf{0.55} & 0.73 & 0.90 & 1.06 & \textbf{17.75}& \textbf{15.99} & \textbf{14.49} & \textbf{12.49} & \textbf{6.41} & \textbf{7.46} & \textbf{7.77} & \textbf{8.17} \\
\midrule
\multirow{3}{*}{Diff.}
& SGMSE+ & 0.37 & 0.53 & 0.71 & 0.88 & 9.81 & 12.35 & 12.08 & 11.10 & \textbf{4.35} & \textbf{7.09} & \textbf{8.60} & \textbf{9.75} \\
& BBED & 0.40 & 0.56 & 0.74 & 0.89 & 18.01 & 16.74 & 15.57 & 13.56 & 3.63 & 5.44 & 6.87 & 7.83 \\
& GALDSE & 0.30 & 0.46 & 0.64 & 0.80 & \bfseries 18.24 &  16.75 &  15.51 &  13.54 & 1.33 & 2.82 & 4.14 & 4.71 \\
& FlowSE& \textbf{0.41}& \textbf{0.57} &\textbf{0.75} & \textbf{0.89}& 18.13& \bfseries 16.80& \underline{\bfseries 15.59} & \underline{\bfseries 13.63} & 3.91 & 5.96 & 7.20 & 8.09 \\
\midrule
\multirow{2}{*}{GAN}
& DisCoGAN & \underline{\bfseries 0.58} &  0.79 & \underline{\bfseries 1.01} & \underline{\bfseries 1.22} & 17.48 & 16.06 & 14.81 & 12.96 &  7.19 &  8.53 &  9.33 &  9.94 \\
& NoCoGAN & 0.51 & 0.71 & 0.91 & 1.10 & 16.85 & 15.51 & 14.20 & 12.42 & 5.63 & 6.84 & 7.63 & 8.16 \\
& NCSN++ (GAN)  & 0.58 &  \underline{\bfseries0.80} & 0.98 & 1.16 & \underline{\textbf{18.54}} & \underline{\textbf{16.97}} & \textbf{15.54} & \textbf{13.61} & \underline{\bfseries 8.58} & \underline{\bfseries 9.90} &  \underline{\bfseries 10.53} & \underline{\bfseries 11.12}  \\
& CMGAN & 0.49 & 0.68 & 0.87 & 1.06 & 17.58 & 16.02 & 14.70 & 12.86 & 7.91 & 9.04 & 10.07 & 10.66  \\

\bottomrule
\end{tabular}
}
\label{tab:lowsnrmatched}
\vspace{-2em}
\end{table*}

\noindent \textbf{Evaluation Metrics and Criteria:} We evaluate the different methods using both intrusive and non-intrusive metrics. Specifically, we use PESQ~\cite{rix2001perceptual}, frequency-weighted segmental SNR (FwSegSNR)~\cite{ma2009objective}, SI-SDR~\cite{le2019sdr}, DNSMOS~\cite{reddy2021dnsmos}, and SCOREQ (reference-based)~\cite{ragano2024scoreq}. To evaluate hallucination effects, we report \ac{WER} and \ac{CER} using the Whisper (base) ASR system~\cite{radford2023robust} with the JiWER toolkit~\cite{morris2004and}, along with the Levenshtein phoneme similarity (LPS)~\cite{pirklbauer2023evaluation}. For computational complexity, we report giga multiply–accumulate operations per second (GMACs) and the number of model parameters. The best candidate  models among different epoch-wise checkpoints for each method are selected based on PESQ and SI-SDR performance on the validation set. 

\begin{table}[t]
\caption{Evaluation of models trained on the high-\ac{SNR} dataset in low-\ac{SNR} scenarios (mismatched condition).}
\centering
\scriptsize
\setlength{\tabcolsep}{3pt} 
\resizebox{\columnwidth}{!}{%
\begin{tabular}{ll ccc ccc}
\toprule
& & \multicolumn{3}{c}{\textbf{$\Delta$PESQ}(↑)} & \multicolumn{3}{c}{\textbf{$\Delta$ SI-SDR (dB)}(↑)} \\
\cmidrule(lr){3-5} \cmidrule(lr){6-8}
\textbf{Method} & \textbf{Model} & \makecell{[-11,-8]} & \makecell{[-7,-4]} & \makecell{[-3,0]} & \makecell{[-11,-8]} & \makecell{[-7,-4]} & \makecell{[-3,0]} \\
\midrule

\multirow{2}{*}{Disc.} 
& NoCoGAN (D) & \textbf{0.73} & \textbf{0.93} & \textbf{1.10} & \textbf{15.20} & 13.75 & 11.84 \\
& NCSN++ (D)  & 0.64 & 0.82 & 0.99 & 14.94 & \textbf{14.02} & \textbf{12.06} \\ 
\midrule

\multirow{4}{*}{Diff.} 
& SGMSE+      & 0.31 & 0.53 & 0.78 & 7.55 & 9.17 & 10.22 \\
& BBED       & 0.39 & 0.54 & 0.66 & 14.66 & 14.07 & 12.51 \\
& GALDSE     & 0.37 & 0.57 & 0.74 & \textbf{15.22} & \textbf{14.58} & \textbf{13.54} \\
& FlowSE     & \textbf{0.45} & \textbf{0.62} & \textbf{0.79} & 15.10 & 14.22 & 12.65 \\
\midrule 

\multirow{3}{*}{GAN} 
& DisCoGAN   & 0.70 & \underline{\bfseries 0.94} & \underline{\bfseries 1.16} & 15.31 & 14.31 & 12.63 \\
& NoCoGAN    & 0.64 & 0.86 & 1.09 & 14.64 & 13.65 & 12.10 \\
& NCSN++ (GAN) & \underline{\bfseries 0.72} & 0.93 & 1.14 & \underline{\bfseries 16.03} & \underline{\bfseries 15.13} & \underline{\bfseries 13.41} \\

\bottomrule
\end{tabular}
}
\label{tab:lowsnrmismatched}
\vspace{-2em}
\end{table}

\subsection{Results and Discussion}

\noindent \textbf{High- and Low-SNR scenarios:}
In Tab.~\ref{tab:DNSnoreverb}, we present the results for matched high-\ac{SNR} scenarios on the DNS Challenge non-reverberant test set. We observe that \ac{GAN}-based methods outperform both diffusion and discriminative methods in terms of PESQ, SI-SDR and SCOREQ improvement. The discriminative methods achieve PESQ scores comparable to \ac{GAN}-based approaches; however, they lag behind in other objective metrics. In terms of the non-intrusive DNN-based metric DNSMOS, diffusion-based methods slightly outperform other approaches, potentially indicating stronger generative capabilities compared to \acp{GAN} and discriminative methods~\cite{shetu2025leveraging}.

Based on the results from high-\ac{SNR} matched scenarios, we trained the best-performing (and distinct) models with the low-\ac{SNR} dataset, as described above.  However, in low-\ac{SNR} scenarios, \ac{GAN}-based methods clearly outperform discriminative and diffusion-based methods under matched and mismatched conditions, as shown in Tabs.~\ref{tab:lowsnrmatched} and~\ref{tab:lowsnrmismatched}. In matched scenarios, diffusion-based methods achieve comparable or better SI-SDR performance to GAN-based methods (e.g., FlowSE achieves SI-SDR improvements of $16.80$, $15.59$, and $13.63$ for the \ac{SNR} groups $[-11, -8]$, $[-7, -4]$, and $[-3, 0]$, respectively). However, they perform poorly in terms of PESQ and FwSegSNR, which may indicate that iterative refinement methods are less suitable for extremely low-\ac{SNR} conditions, potentially due to over-denoising or oscillations while converging toward producing clean speech outputs. It is important to note that NCSN++ (GAN) (i.e., the model architecture used in diffusion-based methods but trained with a GAN objective) performs on par with or better than DisCoGAN in both matched and mismatched scenarios. In low-\ac{SNR} scenarios, NCSN++ (GAN) outperforms other methods by a clear margin in terms of FwSegSNR ($9.90$, $10.53$, and $11.12$) and SI-SDR ($16.03$, $15.13$, and $13.41$ dB) improvement across \ac{SNR} groups $[-11, -8]$, $[-7, -4]$, and $[-3, 0]$, in  both matched and mismatched scenarios, respectively. 

\begin{figure}[t]
\centering
\resizebox{0.95\linewidth}{!}{%
  \input{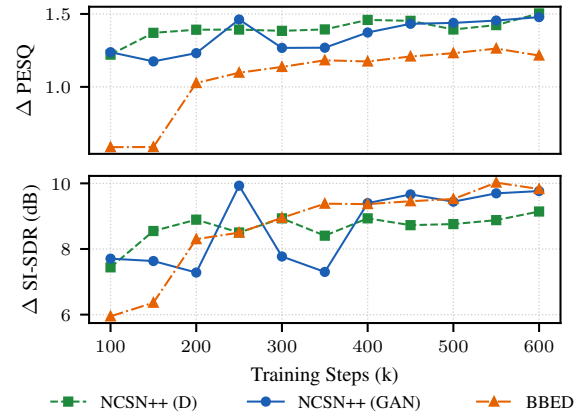}
}
\vspace{-2em}
\caption{Training convergence in terms of PESQ and SI-SDR improvement on the DNS Challenge non-reverb test set.} 
\label{fig:datasteps}
\vspace{-1.5em}
\end{figure}

\begin{figure}[t]
\centering
\resizebox{0.95\linewidth}{!}{%
  \input{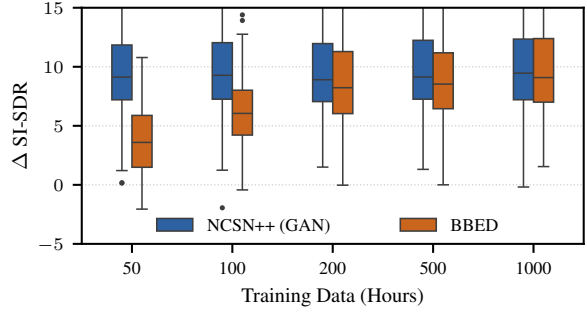}
}
\vspace{-2em}
\caption{SI-SDR improvement on the DNS Challenge non-reverb test set vs training data volume, illustrating the impact of training data.} 
\label{fig:datahours}
\vspace{-1.5em}
\end{figure}

\begin{table}[!t]
\caption{\ac{WER}, \ac{CER}, and LPS results for GAN and diffusion models and the noisy reference. Results are reported in percentage ($\%$).}
\centering
\scriptsize
\setlength{\tabcolsep}{3pt} 
\resizebox{\columnwidth}{!}{%
\begin{tabular}{ll cccccc}
\toprule
& & \multicolumn{2}{c}{\textbf{WER (↓)}} & \multicolumn{2}{c}{\textbf{CER (↓)}} & \multicolumn{2}{c}{\textbf{LPS (↑)}} \\
\cmidrule(lr){3-4} \cmidrule(lr){5-6} \cmidrule(lr){7-8}
\textbf{Method} & \textbf{Model} & \makecell{[-7,-4]} & \makecell{[-3,0]} & \makecell{[-7,-4]} & \makecell{[-3,0]} & \makecell{[-7,-4]} & \makecell{[-3,0]} \\
\midrule

\multirow{1}{*}{Ref.} 
& Noisy        & 89 & 39 & 66& 24 & 55& 69 \\
\midrule

\multirow{3}{*}{GAN}
& NoCoGAN      & 43 & 31 & 28 & 19 & 84 & 91 \\
& DisCoGAN     &  \underline{\bfseries  39} & 26 & \underline{\bfseries  25} & 16 & \underline{\bfseries  85} & \underline{\bfseries 92} \\
& NCSN++ (GAN) & 42 &  \underline{\bfseries  26} & 29 & \underline{\bfseries  13} & \underline{\bfseries  85} & \underline{\bfseries  92} \\
\midrule
\multirow{3}{*}{Diff.}
& BBED         & \bfseries 43 & 39 & 30 & \bfseries 17 & \bfseries 82 & \bfseries 90 \\
& FlowSE       & 46 & \bfseries 39 & \bfseries 28 & 24 & 82 & 89 \\
& GALDSE       & 62 & 32 & 45 & 18 & 81& 89 \\

\bottomrule
\end{tabular}%
}
\label{tab:wer_lps_lsd_with_noisy}
\vspace{-2em}
\end{table}

\noindent \textbf{Training Convergence and Dataset:}
In Fig.~\ref{fig:datasteps}, we analyze the training convergence characteristics of different methods using the same NCSN++ backbone network in terms of PESQ and SI-SDR improvement. We observe that the discriminative model achieves peak performance at around $200$k steps and shows minimal improvement up to $600$k steps. The \ac{GAN}-based NCSN++ (GAN) model reaches peak performance at approximately $250$k steps; however, its performance exhibits noticeable oscillations before stabilizing around $400$k steps. This instability in GAN training can be attributed to the competing optimization of the generator and discriminator. In contrast, the diffusion-based BBED model demonstrates a more stable training behavior, with gradual performance improvements. It required around $300$k steps to achieve SI-SDR performance comparable to the discriminative NCSN++ (D) model and approximately $400$k steps to approach the performance of the \ac{GAN}-based method. However, it did not achieve PESQ improvements comparable to discriminative or \ac{GAN}-based approaches.

To assess the impact of volume of training data, we created four subsets of the high-\ac{SNR} dataset with durations of $50$, $100$, $200$, and $500$ hours by randomly sampling the training data. As shown in Fig.~\ref{fig:datahours}, NCSN++ (GAN) achieves peak performance in terms of SI-SDR with only $50$ hours of training data. In contrast, the diffusion-based BBED method is significantly more sensitive to data scale, requiring at least $200$ hours of data to achieve comparable results to its peak performance. These results suggest that \ac{GAN}-based approaches are more data-efficient and converge faster for \ac{SE} tasks.

\noindent \textbf{Hallucination Effects:}
In Tab.~\ref{tab:wer_lps_lsd_with_noisy}, we report \ac{WER}, \ac{CER}, and LPS to assess the hallucination tendencies of generative methods. All methods consistently improve these metrics, indicating that they appear to introduce only limited hallucination effects within the evaluated SNR ranges. GAN-based methods appear to outperform diffusion-based methods in these metrics. In particular, NCSN++ (GAN) and DisCoGAN achieve higher phoneme similarity scores of $85\%$ and $92\%$ in the SNR groups $[-7, -4]$ and $[-3, 0]$, respectively.  However, our inspections also revealed that in very low \ac{SNR} scenarios (e.g., below $-7$~dB), these metrics degrade significantly. In this \ac{SNR} range, we also observed spurious spectral content in the enhanced spectrograms of the generative methods which was not present in the original clean speech signal. These results suggest that conditional generative models may rely heavily on the noisy input signal $\mathbf{y}$, leading to limited hallucination under moderate to high \ac{SNR} matched conditions; however, in very low \ac{SNR} scenarios, where the noisy signal $\mathbf{y}$ may be almost entirely masked by noise, the generative methods tend to hallucinate more \cite{shetu2025leveraging}.

\begin{figure}[!t]
\centering
\includegraphics[width=0.9\linewidth]{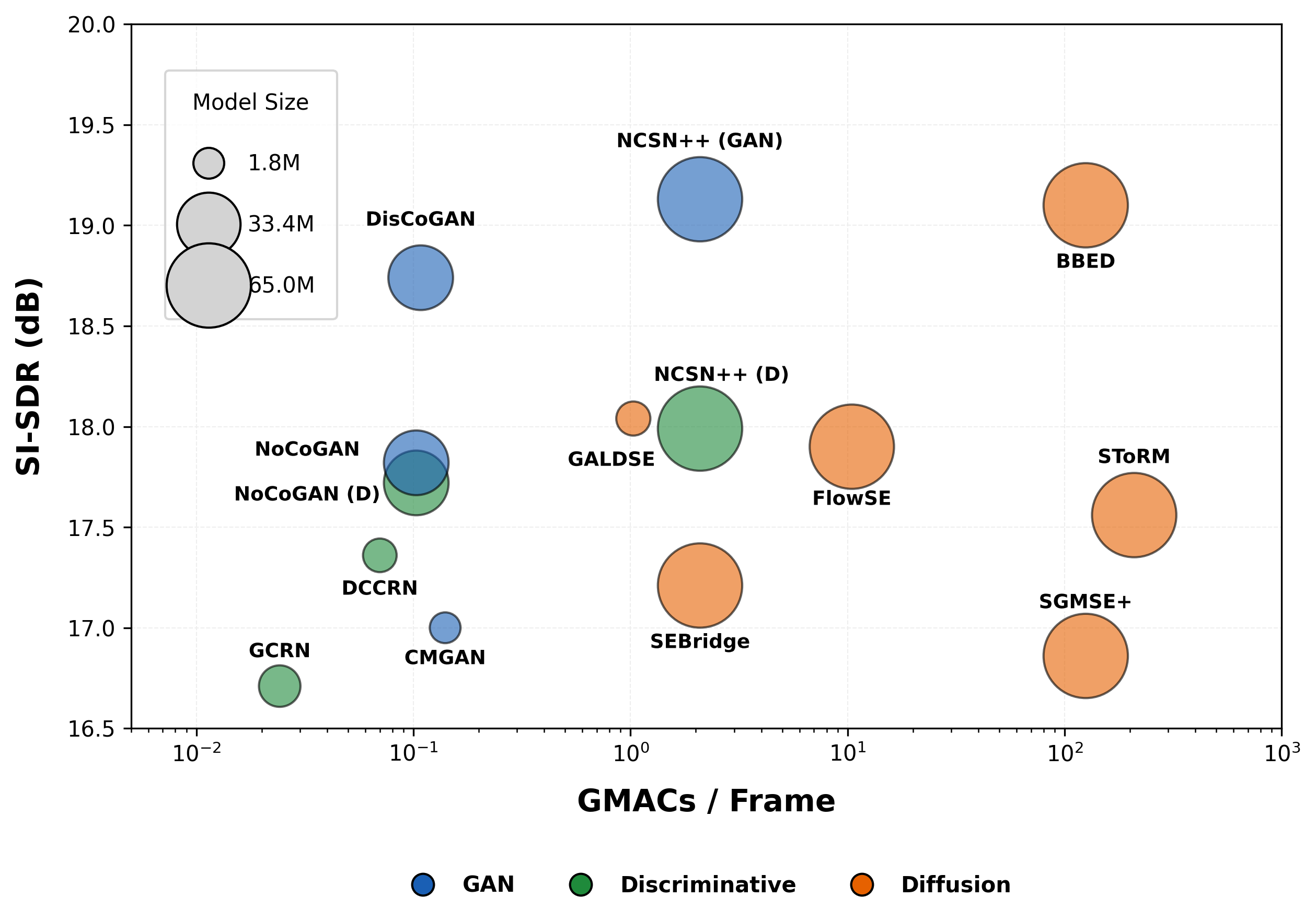}
\vspace{-1.5em}
\caption{Comparison of model complexity across different methods in terms of GMACs and number of parameters.}
\label{fig:lis2}
\vspace{-1.5em}
\end{figure}

\noindent \textbf{Complexity-Performance Tradeoff:}
In Fig.~\ref{fig:lis2}, we show the complexity of different \ac{SE} methods in terms of GMACs and the number of parameters. We observe that most of the evaluated diffusion-based methods are significantly more complex, both in terms of GMACs and parameter count. SGMSE+ (30 steps) and SToRM (50 steps)  are at least $60-100\times$ more computationally expensive in terms of GMACs than NCSN++ (GAN) and NoCoGAN, mainly due to the large number of network evaluations during the reverse diffusion process. FlowSE and SEBridge reduce the number of network evaluations to a single or a few steps (e.g., five); however, they still remain considerably complex due to the use of the computationally expensive NCSN++ backbone. In contrast, discriminative and GAN-based methods relying on single-step inference, show less computational complexity than diffusion-based methods. NCSN++, which is computationally more intensive than NoCoGAN and DisCoGAN, achieves better results in several objective metrics in different scenarios for both GAN-based and discriminative training methods, which indicates that higher model complexity, also scales to better objective performance for discriminative and GAN-based methods. 

\noindent \textbf{Discussion:} Our experimental results suggest that, at least for \ac{SE} tasks, which benefit from the strong conditioning provided by the noisy signal $\mathbf{y}$, very large models or multiple network evaluations, as required in diffusion-type methods, are not always necessary. In high-SNR scenarios, the discriminative model remains competitive to generative approaches while showing significant benefits in computational efficiency \cite{serbest2025deepfiltergan,shetu2023ultra,schroter2022deepfilternet2,shetu2024hybrid}. We hypothesize that generative methods, particularly diffusion-type approaches, may be more beneficial than discriminative methods for tasks that are inherently generative in nature, such as speech reconstruction/synthesis \cite{huang2022fastdiff} and bandwidth extension \cite{moliner2024blind}. Under the evaluated conditions, diffusion-based methods do not demonstrate clear advantages in terms of the complexity-performance trade-off.

\section{Conclusions}
In this study, we show that, while generative methods improve \ac{SE} performance, particularly in low-SNR scenarios and mismatched conditions, the complexity–performance trade-off does not always justify the performance gains, especially for diffusion-based methods. Our analysis also shows that, for \ac{SE} tasks, discriminative and GAN-based methods result in faster training times and can achieve better efficiency in terms of training data. Future studies should focus on other \ac{SE}-related tasks and on the impact of different neural network architectures across training paradigms.


\let\oldbibliography\thebibliography
\renewcommand{\thebibliography}[1]{%
  \oldbibliography{#1}%
  \footnotesize 
  \setlength{\itemsep}{-0.0ex} 
  \setlength{\parsep}{0pt}  
  \setlength{\parskip}{0pt} 
  \setlength{\leftmargin}{1em} 
}

\bibliographystyle{IEEEbib}
\bibliography{strings,refs}

\end{document}